\documentclass[%
 reprint,
 amsmath,amssymb,
 aps,
 prd,
 superscriptaddress,
 nofootinbib,
 eqsecnum
]{revtex4-2}
\usepackage{svg}
\usepackage{graphicx}%
\usepackage{mathtools}
\usepackage{epstopdf}
\usepackage{comment}
\usepackage{bm}%
\usepackage{ifthen}
\usepackage{tensor}
\usepackage{hyperref}%
\hypersetup{
    colorlinks,
    linkcolor = {blue!100!black},
    citecolor = {blue!100!black},
    urlcolor = {blue!100!black}
}
\usepackage{adjustbox}
\usepackage{physics}

\newcommand{\beq}{\begin{equation}}
\newcommand{\eeq}{\end{equation}}
\newcommand{\beqa}{\begin{eqnarray}}
\newcommand{\eeqa}{\end{eqnarray}}
\newcommand{\bite}{\begin{itemize}}
\newcommand{\eat}{\end{itemize}}
\newcommand{\rmit}{\affiliation{Centre for Quantum Computation and Communication Technology, School of Science, RMIT University, Melbourne, VIC 3000, Australia}}
\newcommand{\rmitsci}{\affiliation{Discipline of Physics, School of Science, RMIT University, Melbourne, VIC 3000, Australia}}

\let\originalleft\left
\let\originalright\right
\renewcommand{\left}{\mathopen{}\mathclose\bgroup\originalleft}
\renewcommand{\right}{\aftergroup\egroup\originalright}

\newcommand{\kcomp}[1][]{
\ifthenelse{ \equal {#1} {} }{k_{\mathrm{C}}}
{k_{\mathrm{C},{\mathrm{#1}}}}
}
\newcommand{\lcomp}{\lambda_{\mathrm{C}}}
\newcommand{\sonic}{{\mathrm{s}}}
\newcommand{\internal}{{\mathrm{in}}}
\newcommand{\external}{{\mathrm{ex}}}
\newcommand{\particle}{{\mathrm{p}}}
\newcommand{\observer}{{\mathrm{o}}}
\newcommand{\milk}{{\mathrm{M}}}
\newcommand{\honey}{{\mathrm{H}}}
\newcommand{\laboratory}{{\mathrm{L}}}

\newcommand{\pd}[2]{\frac {\partial {#1}} {\partial {#2}}}

\DeclareMathOperator{\diag}{diag}

\allowdisplaybreaks[4]

\begin{document}

\title{Tachyonic media in analogue models of special relativity}

\author{Sundance O. Bilson-Thompson}%
 \email{sundance.bilson-thompson@adelaide.edu.au}
\rmit
\affiliation{%
 School of Physics, Chemistry and Earth Sciences, University of Adelaide, SA 5005 Australia
}%

\author{Scott L. Todd}
 \email{scott@todd.science}
\rmitsci

\author{James~Read}
\email{james.read@philosophy.ox.ac.uk}
\affiliation{Faculty of Philosophy, University of Oxford, OX2 6GG, United Kingdom}

\author{Valentina Baccetti}
 \email{valentina.baccetti@rmit.edu.au}
\rmitsci

\author{Nicolas~C. Menicucci}
 \email{nicolas.menicucci@rmit.edu.au}
\rmit

\date{\today}%

\begin{abstract} 
In sonic models of special relativity, 
the fact that the sonic medium violates (ordinary) Lorentz symmetry is apparent to observers external to the sonic medium but not to a class of observers existing within the medium itself. 
We show that the situation is symmetric:
internal observers will judge physics in the external laboratory to violate their own sonic Lorentz symmetries. We therefore treat all observers on an equal footing such that each is able to retain a commitment to their own Lorentz symmetries. We then generalize beyond the case of subsystem-environment decompositions to situations in which there exist multiple phonon fields, all obeying Lorentz symmetries but with different invariant speeds. In such cases, we argue that all observers have freedom to choose which field is symmetry preserving, and so---in a certain precise sense---which other fields are perceived as having an `ether.' This choice is influenced---but not determined---by a desire for simplicity in the description of physical laws.
Sending information faster than sound serves as a model of tachyonic signalling to a distant receiver. Immutable causality of the laboratory setup when perceived externally to a sonic medium manifests internally through the confinement of the tachyons to an apparent ether (with a rest frame), which we call a ``tachyonic medium,'' thereby preventing tachyonic exchange from emulating the scenario of a round-trip signal travelling into an observer's past causal cone. The assignment of sonic-Lorentz-violating effects to fields that obey `photonic' Lorentz symmetries thus ensures that causality associated with the `sonic' Lorentz symmetries is preserved.
\end{abstract}

\maketitle

\section{Introduction}
\label{sec:intro}

It is well-known that distinct physical systems can behave in similar ways. Consider, for example, the fact that the same mathematics can be used to model an oscillating spring as an LC circuit~\cite{Fletcher}. In this article, we further explore the fact---already investigated in recent works such as~\cite{todd2020analogue, TM, todd2021particle, ChengRead}---that the mathematical structure of Minkowski spacetime does not depend upon the exchange of light signals specifically but rather that many distinct physical systems and phenomena manifest Lorentz symmetries, albeit with potentially different invariant speeds (e.g.,~the speed of light versus speed of sound).

Relativistic analogies---that is, models that are explicitly non-relativistic but that nonetheless capture some of the phenomenology of relativistic systems---have been well-explored in the literature in recent years, though usually in the context of general relativity. Such models are referred to as \emph{analogue gravity models},%
\footnote{`Analogue' in the sense of analogous/analogy.}
and are often considered to have originated in the work of Unruh~\cite{unruhExperimentalBlackHoleEvaporation1981} (though earlier models, such as that of Gordon~\cite{gordonZurLichtfortpflanzungNach1923}, can rightly be described as analogue gravity models). The review article by Barcel\'{o} et al.~\cite{barceloAnalogueGravity2011} provides a thorough overview of the analogue gravity research endeavour as of 2011, and some of the important developments between 2011 and 2020 are discussed by Jacquet et al.~\cite{jacquetNextGenerationAnalogue2020}

While analogue gravity models are interesting---and potentially of great utility---in their own right, they are not explicitly the focus of our investigations. %
We wish to understand which aspects of relativistic physics are in principle emergent (and thus, in principle, could be captured by an analogue model) and which aspects of relativistic physics are fundamental (and thus, cannot---even in principle---be captured by an analogue model). In essence, we wish to understand how far analogue models of relativistic physics (such as analogue-gravity models) can be pushed. The answer to such a question is not always obvious. For example, one recent study concludes that, despite the laboratory physics displaying an underlying Minkowskian (or Galilean) causal structure, simulation of some spacetimes with closed-timelike curves is possible in analogue models~\cite{barceloChronologyProtectionImplementation2022}.

This current work is preceded by Refs.~\cite{TM, todd2020analogue}, in which the behaviour of sonically relativistic systems was analysed through reference to an external environment (the lab)---in which information can propagate at the speed of light---and an acoustic subsystem (the sonic medium) consisting of a solid or fluid. Internal to the acoustic subsystem, sound pulses can be thought to take the place of light signals, and thus, experiments analogous to those performed with light in special relativity can be performed with sound pulses. For example, the passage of time can be measured using \emph{sound clocks}, and thus the effects of relative motion on temporal measurements can be investigated. We refer to observers embedded within an acoustic subsystem as \emph{internal observers} and ordinary observers in the lab as \emph{external observers}. It was shown in Ref.~\cite{TM} that internal observers who perform experiments that measure lengths and durations via the exchange of phonons\footnote{In Ref.~\cite{TM}, internal observers were dubbed \emph{in-universe observers}.} will %
observe length contraction and time dilation as described by the usual formulae from special relativity in a spacetime in which Lorentz invariance is obeyed but with the speed of sound $c_\sonic$ replacing the speed of light $c$ in the Lorentz factor---hence,
\begin{equation}
 \gamma_\sonic = \frac{1}{\sqrt{1-{\frac{v^2}{c_\sonic^2}}}}.
\end{equation}
Particle scattering from the perspective of such observers was investigated in~\cite{todd2021particle}, wherein a toy model of scattering between phonons and two different types of particle was considered: the first type of particle obeyed a sonically relativistic energy-momentum relation, whereas the second type of particle was described by a Newtonian energy-momentum relation. While not unexpected, it was demonstrated that---within the constraints of the toy model---the scattering profile of phonons from sonically relativistic particles is insensitive to the absolute state of motion of the experimental apparatus with respect to the medium itself. %
Conversely, 
the scattering profile of phonons from Newtonian (i.e.,~sonic-Lorentz-violating) particles is sensitive to the absolute state of motion with respect to the experimental apparatus and thus---in principle---can reveal information about the medium's rest frame.%
\footnote{As we will show 
throughout the rest of this article, the use here of the qualifier `in principle' can be viewed as the starting point of this current investigation. We will demonstrate that internal observers can, in fact, draw a different 
conclusion from experiments involving sonic-Lorentz-violating phenomena.}
Note that in Ref.~\cite{todd2021particle}, the actual physical mechanism behind the scattering interaction was not considered explicitly, and the nature of the two types of particle under consideration was left unspecified beyond their respective energy-momentum relations. 
Other works exist in the literature, however, regarding particle scattering in analogue models---one such work being the article by Fischer and Visser in which an explicit model is constructed for the particles from which phonons are scattered~\cite{fischerRiemannianGeometryIrrotational2002}.

Our goal in the remainder of this article is to further explore physics from the point of view of observers internal to the sonic medium and observers external to it. In Sec.~\ref{sec:sonics}, we define a class of 
internal observers
who
measure the speed of sound in their medium to be an invariant, and thus whose measurements of lengths and duration respect a sonic Lorentz symmetry. In Secs.~\ref{sec:lagrangians}--\ref{sec:observers}, we discuss how a given physical system (in terms of Lagrangians) might be described from internal and external perspectives. In Secs.~\ref{sec:observers} and~\ref{sec:relofrel}, we argue that the descriptions from each of these settings should be considered as being on an equal footing, and we consider the ramifications of this. In Secs.~\ref{sec:retrocausal} and~\ref{sec:tachyonicmedia}, we raise and resolve some concerns regarding causal paradoxes and (the analogue of) superluminal signalling presented by the preceding work in Refs.~\cite{TM, todd2020analogue}. In so doing, we introduce the concept of a tachyonic medium as a phenomenon seen by {internal observers}. We close in Sec.~\ref{sec:conclusion}.

\section{internal observers}
\label{sec:sonics}
We mentioned above that observers internal to a given medium will regard the speed of sound in that medium as an invariant, leading to time dilation and length contraction effects (and thus, all of the kinematic behaviour associated with special relativity). The observers we wish to consider here will consequently be unaware of the existence of a preferred rest frame for their own medium---and hence the existence of the medium itself---and will conceive of their medium as a special-relativistic Minkowski spacetime with physical laws based upon their own sonic Lorentz transformations (Cf.~Ref.~{\cite[p.~174]{Liberati}}). The reader may already object at this point on the grounds that an observer can perform a Michelson-Morley experiment to detect that the signals they send and receive are supported by a medium. This is true for some classes of observers, but not all~\cite{ChengRead}. We therefore introduce the following three  categories of
observers for consideration:
\begin{enumerate}
    \item observers composed of ordinary fermionic matter (as would exist in the laboratory), using clocks and rulers of fixed scale, also made of the same fermionic matter;
    \item observers restricted to using sound clocks and rulers composed of fermionic matter, but dynamically establishing the scale between their sound clocks and rulers using sound pulses, so as to keep clocks synchronized within their own rest frame (for an explicit illustration of this case, see Ref.~\cite{TM}); and
    \item observers restricted to using clocks and rulers composed of quasiparticle excitations of the sonic medium---or those of another medium at rest with respect to the first one and having the same speed of sound (see Ref.~\cite{barcelo2008real}).   
\end{enumerate}
A crucial part of the Michelson-Morley experiment involves rotating the apparatus to ensure that its dimensions and initial orientation did not accidentally cancel out any evidence of motion through the `ether'. This will indeed be effective for observers of type~1. However, if the observers and their apparatus are composed of quasiparticles---i.e.,~they are of type~3 above---such that the `acoustic atoms' of which they are composed are bound together by acoustic analogues of electrostatic forces (and themselves have a sonic-relativistic dispersion relation), then any direction-dependent modification of the rate at which sounds propagate will be impossible to detect~\cite{barcelo2008real}. This %
echoes the idea of Heaviside that the equipotential surfaces of charges moving through an ether would be distorted into ellipsoids, for in this case there really is a medium to support the forces holding acoustic atoms together to form (macroscopic) acoustic objects~\cite{bell1995teach}. Similarly, for observers of type~2, dynamically establishing length scales by the exchange of phonons will ensure that rotating their apparatus does not detect any direction-dependent effect and, hence, will not detect the existence of a preferred rest frame within that medium~\cite{TM}. Henceforth, we shall restrict our attention to observers of types~2 and~3 and refer to such observers as `internal observers' to a given sonic medium. 

For the sake of brevity and generality when talking about the speed of light, or the speed of sound in different media, we will refer to the \emph{Lorentz-invariant speed}~(LIS) associated with a particular Lorentz group. %
 The usual Lorentz symmetry of the Standard Model has a LIS of~$c$ (the speed of light), while internal observers will perceive a LIS of~$c_\sonic$ (the speed of sound).

In ordinary physics, the LIS also plays the role of a universal upper speed limit. We choose not to emphasize this interpretation because the {internal observers}' LIS of $c_\sonic$ will not necessarily be an upper limit on the speed of particles or excitations in different media or the lab since these may obey Lorentz symmetries with LISs different from $c_\sonic$. (This point is also made in Ref.~\cite[\S5.4]{ChengRead}.)
Note, however, that an object travelling faster or slower than a given LIS will remain as such, respectively, to all observers with that same LIS. That is to say, if some observer with a given LIS measures an object to have a velocity {higher or lower} than their own LIS, then all other observers with the same LIS will also measure its velocity to be {higher or lower, respectively,} than the LIS (though not necessarily with the same value, due to the Lorentz transformation of observers' coordinates). This point is essentially made in Ref.~\cite{Benford}, in which it is demonstrated on geometric grounds that ``[t]he tachyon beam can take on any velocity \emph{outside} the interval from $-c$ to $+c$ relative to a suitably chosen frame.'' It should be borne in mind that this is a consequence of the {internal observer} boosting between different reference frames, not a restriction on how a particle external to the sonic medium may change its motion.

\section{Lagrangians from internal and external perspectives}
\label{sec:lagrangians}

To begin, we will consider a given particle as witnessed by both internal observers (i.e.,~those native to the medium; discussed above) and by external observers (i.e.,~those in the laboratory). Later, we will consider separate particles in two different acoustic media, each obeying a distinct LIS.

\subsection{External particle described by an external observer}
\label{subsec:ex_from_ex}

The Lagrangian for an \emph{external particle}---i.e.,~one obeying regular relativity (LIS of $c$), as viewed by an external observer is simply
\begin{equation}
\mathcal{L}_\external
=
\frac{1}{2}
  \left( 
  \eta^{\mu\nu}\partial_\mu\phi\partial_\nu\phi + \frac{m^2c^2}{\hbar^2}\phi^2
  \right),
  \label{eq:orig_L_int_with_c}
\end{equation}
where $x^{\mu} = (c t, x, y, z)$ are the usual laboratory coordinates, which we denote by standard Greek indices, and we use the ordinary Minkowski inverse metric ${\eta\indices{^\mu^\nu} = \operatorname{diag}(-1,+1,+1,+1)}$, when written in these coordinates, where $\diag$ indicates a diagonal matrix. To be clear, the subscript `ex' indicates the particle type, not the observer describing it. This labelling will be important when we ask how internal observers would view the same particle.

Notice that the coefficient of $\phi^2$ explicitly includes the appropriate fundamental constants, revealing it to be the square of the Compton wavenumber for the particle,
\begin{equation}
    \kcomp
\coloneqq
    \frac{2\pi}{\lcomp}
=
    \frac{mc}{\hbar},
\end{equation}
where $\lcomp=h/m c$ is the Compton wavelength associated with the excitations of the field~$\phi$. Thus, what is usually called the `mass term' is correctly interpreted as a term~$\kcomp^2 \phi^2$ that defines a characteristic length~$\lcomp$ associated with the field, rather than a characteristic mass. (In fact, this is what is always meant by `$m^2 \phi^2$' in quantum field theory once the attendant factor of~$c^2/\hbar^2$ is made explicit.) Since $\kcomp$ has units of inverse distance, both terms in~$\mathcal L$ have the same units, which is a good sanity check.

We choose to write Lagrangians using $\kcomp$ instead of $mc/\hbar$ because different masses may be extracted from the same $\kcomp$ depending upon whether one uses $c$ to relate it to mass (as above) or whether one uses $c_\sonic$ in this role instead. While the former is the right choice for an external observer, internal observers have good reason to use the latter~\cite{todd2021particle}.

As such, we have
\begin{equation}
\mathcal{L}_\external = \frac{1}{2} 
  \left( 
\eta^{\mu\nu}\partial_\mu\phi\partial_\nu\phi + \kcomp[\external]^2\phi^2
  \right)
  \label{eq:L_ex_from_ex}
\end{equation}
for an external particle as viewed by an external observer.
\subsection{Internal particle described by an internal observer}
\label{subsec:in_from_in}

We model \emph{internal particles} as phonons within a material; these are so-named because they are the natural particles that would be observed by internal observers. In crafting our model of a sound-carrying material (i.e.,~a sonic medium), we restrict ourselves to an idealization of an ordinary material that might be present in a laboratory. As such, we may use ordinary Newtonian physics to describe its dynamics. Despite our explicit eschewing of relativity (with LIS of~$c$), the Lagrangian for a multidimensional harmonic lattice, when extended to the continuum, takes a relativistic form with LIS of~$c_\sonic$~\cite{lancaster2014quantum}:
\begin{align}
\mathcal{L}_\internal
&
=
\frac{1}{2} 
\left[
(\vec{\nabla}\phi)^{2}-\frac{1}{c_\sonic^2}
\left(
\pd \phi t
\right)^{2}+ \kcomp[\internal]^2\phi^2 
  \right]
\nonumber \\
&
=
\frac{1}{2} 
  \left[
(\eta_\sonic)\indices{^{ab}}\partial_a\phi\partial_b\phi + \kcomp[\internal]^2\phi^2
  \right],
  \label{eq:L_in_from_in}
\end{align}
where the second line expresses it compactly in the internal observers' coordinates~\cite{TM}, 
${x^a = (c_\sonic t, x, y, z)}$, which we distinguish from those of the lab by using Latin indices%
, and where the sonic inverse metric tensor is~$(\eta_\sonic)\indices{^a^b}$, which takes on the form $\operatorname{diag}(-1, +1, +1, +1)$ when written in these coordinates.%
\footnote{A brief note on subscripts: We use the subscript~`$\sonic$' to label kinematical quantities related to measurements, while we use the subscripts `$\external$' and~`$\internal$' to label quantities related to the type of particle being modelled. We generalize this notation below. We omit subscripts (`$\internal$' or `$\external$') on~$\phi$ for conciseness, with the subscript on~$\mathcal L$ determining its nature.}
In keeping with our previous choice of notation, the subscript `in' indicates the particle type, not the observer describing it.

For the moment, we have chosen the internal and external observers' coordinates to be the same except with the time coordinate rescaled, although we will relax that assumption below. The subscript~`$\sonic$' on the metric expresses the important fact that $\eta\indices{_\mu_\nu}$ and $(\eta_\sonic)\indices{_a_b}$ are not the same object despite each having the same components when written in its natural coordinates. 
The reader should also keep in mind that the coordinates
$x^a$ and $x^\mu$ are not related by a Lorentz transformation. These observations will become important when we start to explore how each type of particle is viewed by the other type of observer.

\subsection{General framework for describing particles by external and internal observers}
\label{subsec:generalframework}
Although Eq.~\eqref{eq:L_ex_from_ex} is manifestly Lorentz invariant with a LIS of~$c$, our thesis throughout this work is that internal observers may have occasion to re-express this equation with respect to an entirely different Lorentz symmetry---specifically, one with a LIS of~$c_\sonic$. We will discuss why these observers may wish to do so later on, but for now, we will simply show that this can be done at the cost of introducing a (sonic-)Lorentz-violating term. 

To proceed, we invoke a coordinate transformation~$K\indices{^a_\mu}$ that expresses the internal coordinates in terms of the external coordinates, as well as its inverse~$\bar K\indices{^\mu_a}$, which does the reverse.%
\footnote{Note that $K$ is not a Lorentz transformation since it is relating coordinates native to different Lorentz symmetries, so we need a different object, $\bar K$, for the inverse transformation. That is, $\bar K\indices{^\mu_a} \neq K\indices{_a^\mu}$, which means the inverse matrix cannot be obtained by a transpose of the original matrix.}
We shall at first consider the simple case where the sonic medium is at rest with respect to external (laboratory) coordinates~$x^\mu$. However, the formalism we develop can be generalized to the case where the medium is in linear motion with respect to the laboratory (but not rotating). We write the coordinate transformations as
\begin{align}
\label{eq:xtrans}
    x^a &= K\indices{^a_\mu} x^\mu,
&
    x^\mu &= \bar K\indices{^\mu_a} x^a,
\\
\label{eq:partialtrans}
    \partial\indices{_a}
&=
    \bar K\indices{^\mu_a} \partial\indices{_\mu}
    ,
&
   \partial\indices{_\mu}
&=
    K\indices{^a_\mu} \partial\indices{_a}
    ,
\\
\label{eq:Kdelta}
    K\indices{^a_\mu} \bar K\indices{^\mu_b} &= \delta\indices{^a_b},
&
    \bar K\indices{^\mu_a} K\indices{^a_\nu} &= \delta\indices{^\mu_\nu}
    .
\end{align}
Note that we used the chain rule on Eqs.~\eqref{eq:xtrans} to obtain the transformation rules for $\partial\indices{_\mu}$ and $\partial\indices{_a}$, Eqs.~\eqref{eq:partialtrans}.
Eqs.~\eqref{eq:xtrans} and~\eqref{eq:partialtrans} show the transformation rules for upper and lower indices, respectively, for all objects so that contractions remain valid.%
\footnote{We defer formulating the full mathematical formalism for working with two different Lorentz symmetries to future work. Here we present only the relations required. (For instance, the rules for raising and lowering indices are subtle and introduce complications not needed here.)}

At this point, we can see why a distinction was made between $\eta\indices{^\mu^\nu}$ and~$(\eta_\sonic)\indices{^a^b}$. Specifically, we can now write each inverse metric tensor in the coordinates of the other Lorentz symmetry:
\begin{align}
\label{eq:eta_ext_from_int}
    \eta\indices{^\mu^\nu}
    K\indices{^a_\mu}
    K\indices{^b_\nu}
&\eqqcolon
    \eta\indices{^a^b},
\end{align}
which is the expression of the external inverse metric in the coordinates of the internal observers. Analogously,
\begin{align}
\label{eq:eta_int_from_ext}
    (\eta_\sonic)\indices{^a^b} 
    \bar K\indices{^\mu_a}
    \bar K\indices{^\nu_b}
&\eqqcolon
    (\eta_\sonic)\indices{^\mu^\nu}
\end{align}
expresses the internal inverse metric in external coordinates.

The Lagrangian in Eq.~\eqref{eq:L_ex_from_ex} may be re-expressed in terms of $c_\sonic$
to yield a Lagrangian for an external particle viewed by an internal observer, as follows:
\begin{align}
\label{eq:L_ex_from_in_general}
    \mathcal{L}_\external
&=
    \frac{1}{2} 
    \left(
    \eta\indices{^\mu^\nu}
    \partial\indices{_\mu} \phi
    \partial\indices{_\nu} \phi
    +
    \kcomp[\external]^2
    \phi^2
    \right)
\nonumber \\
&=
    \frac{1}{2} 
    \left(
    \bar K\indices{^\mu_a}
    \bar K\indices{^\nu_b}
    \eta\indices{^a^b}
    K\indices{^c_\mu}
    K\indices{^d_\nu}
    \partial\indices{_c} \phi
    \partial\indices{_d} \phi
    +
    \kcomp[\external]^2
    \phi^2
    \right)
\nonumber \\
&=
    \frac{1}{2} 
    \left(
    \delta\indices{_a^c}
    \delta\indices{_b^d}
    \eta\indices{^a^b}
    \partial\indices{_c} \phi
    \partial\indices{_d} \phi
    +
    \kcomp[\external]^2
    \phi^2
    \right)
\nonumber \\
&=
    \frac{1}{2} 
    \left(
    \eta\indices{^a^b}
    \partial\indices{_a} \phi
    \partial\indices{_b} \phi
    +
    \kcomp[\external]^2
    \phi^2
    \right)
    ,
\nonumber \\
\intertext{after which we extract the sonic metric,}
&=
    \frac{1}{2} 
    \left[
    (\eta + \eta_\sonic - \eta_\sonic)
    \indices{^a^b}
    \partial\indices{_a} \phi
    \partial\indices{_b} \phi
    +
    \kcomp[\external]^2
    \phi^2
    \right]
    ,
\nonumber \\
\intertext{and write it in the final form:}
    \mathcal{L}_\external
&=
    \frac{1}{2} 
    \left[
    (\eta_\sonic)
    \indices{^a^b}
    \partial\indices{_a} \phi
    \partial\indices{_b} \phi
    +
    \kcomp[\external]^2
    \phi^2
    +
    (\eta - \eta_\sonic)
    \indices{^a^b}
    \partial\indices{_a} \phi
    \partial\indices{_b} \phi
    \right]
    .
\end{align}
Analogously, the Lagrangian for an internal particle viewed by an external observer is found to be
\begin{align}
\label{eq:L_in_from_ex_general}
    \mathcal{L}_\internal
&=
    \frac{1}{2} 
    \left[
    \eta
    \indices{^\mu^\nu}
    \partial\indices{_\mu} \phi
    \partial\indices{_\nu} \phi
    +
    \kcomp[\internal]^2
    \phi^2
    +
    (\eta_\sonic - \eta)
    \indices{^\mu^\nu}
    \partial\indices{_\mu} \phi
    \partial\indices{_\nu} \phi
    \right]
    .
\end{align}
The first two terms of Eqs.~\eqref{eq:L_ex_from_in_general} and~\eqref{eq:L_in_from_ex_general} remain invariant under Lorentz transformations native to the observers (internal and external, respectively), while the third term breaks Lorentz invariance due to the presence of the foreign metric tensor ($\eta$ and $\eta_\sonic$, respectively).

Recall that we are currently working in the special case where the sonic medium is at rest with respect to the laboratory coordinates. Thus the internal and external coordinate systems share spatial coordinates and the direction of time but with a different scaling:
\begin{subequations}
\label{eq:standard_coords}
\begin{align}
    x\indices{^\mu}
&=
    (c t, x, y, z),
\\*
    x\indices{^a}
&=
    (c_\sonic t, x, y, z).
\end{align}
\end{subequations}
Quite simply, then,
\begin{subequations}
\label{eq:Kdef}
\begin{align}
    K\indices{^a_\mu} = \diag(c_\sonic/c, 1, 1, 1),
\\*
    \bar K\indices{^\mu_a} = \diag(c/c_\sonic, 1, 1, 1).
\end{align}
\end{subequations}
In these coordinates, the (inverse) metric tensors $\eta$ and $\eta_\sonic$ have the standard components in their respective coordinates,
\begin{subequations}
\label{eq:eta_inv_own_coords}
\begin{align}
    \eta\indices{^\mu^\nu}
&=
    \diag(-1,1,1,1),
\\
    (\eta_\sonic)\indices{^a^b}
&=
    \diag(-1,1,1,1),
\end{align}
\end{subequations}
but in each others' coordinates (Eqs.~\eqref{eq:eta_ext_from_int} and~\eqref{eq:eta_int_from_ext}), they take a different form:
\begin{subequations}
\label{eq:eta_inv_other_coords}
\begin{align}
    \eta\indices{^a^b}
&=
    \diag(-c_\sonic^2/c^2,1,1,1),
\\
    (\eta_\sonic)\indices{^\mu^\nu}
&=
    \diag(-c^2/c_\sonic^2,1,1,1).
\end{align}
\end{subequations}
Thus,
\begin{subequations}
\label{eq:eta_inv_diff}
\begin{align}
    (\eta - \eta_\sonic)
    \indices{^a^b}
&=
    \diag(\gamma(c_\sonic)^{-2},0,0,0),
\\
    (\eta_\sonic - \eta)
    \indices{^\mu^\nu}
&=
    \diag(\gamma_\sonic(c)^{-2},0,0,0).
\end{align}
\end{subequations}
where
\begin{align}
    \gamma(v) =
    \left(
    1 - \frac {v^2} {c^2}
    \right)^{-1/2}
\end{align}
is the ordinary Lorentz factor with respect to~$c$, and 
\begin{align}
    \gamma_\sonic(v) =
    \left(
    1 - \frac {v^2} {c_\sonic^2}
    \right)^{-1/2}
\end{align}
is the ``sonic Lorentz factor'' with respect to~$c_\sonic$~\cite{TM}.

In order to keep the presentation general, we can write the inverse metric differences as proportional to
dyadic products of the terms 
\begin{align}
     (f_\external)\indices{^a}
&=
    (1,0,0,0)
&
&\text{and}
&
    (f_\internal)\indices{^\mu}
&=
    (1,0,0,0)
\label{eq:f_four_velocities}
\end{align}
such that
\begin{subequations}
\label{eq:eta_inv_diff_dyadic_prod}
\begin{align}
    (\eta - \eta_\sonic)
    \indices{^a^b}
&=
    \gamma(c_\sonic)^{-2}
    (f_\external)\indices{^a}
    (f_\external)\indices{^b}
    ,
\\
    (\eta_\sonic - \eta)
    \indices{^\mu^\nu}
&=
    \gamma_\sonic(c)^{-2}
    (f_\internal)\indices{^\mu}
    (f_\internal)\indices{^\nu}.
\end{align}
\end{subequations}
Although we will not consider the case in which the acoustic medium is in motion relative to the laboratory coordinates, $(f_\external)\indices{^a}$ and $(f_\internal)\indices{^\mu}$ can be shown to transform as four-vectors under their respective Lorentz boosts. We therefore interpret these terms as four-velocities (sonic and ordinary, respectively) that indicate the relative state of motion of the acoustic medium and the lab. The simple form in eq.~(\ref{eq:f_four_velocities}) applies when the medium and lab are mutually at 
rest.\footnote{Formulae of this kind are already known in the literature: see e.g.,~Ref.~\cite{MenonRead}; note also that the formula relating a Newtonian spatial metric to a Minkowski metric via a timelike vector field discussed in Ref.~\cite{Barrett} is a special case of the above in which the lightcones of one of the metrics under consideration are widened completely.}
Although they have the same components in these particular coordinates, it should be kept in mind that these are distinct four-vectors, and each respects a different Lorentz symmetry.

The final forms of our Lagrangians %
are
\begin{align}
\label{eq:L_ex_from_in_general_with_vel}
   \mathcal{L}_\external
&=
    \frac{1}{2} 
    \left[
    (\eta_\sonic)
    \indices{^a^b}
    \partial\indices{_a} \phi
    \partial\indices{_b} \phi
    +
    \kcomp[\external]^2
    \phi^2
    +
    \gamma(c_\sonic)^{-2}
    [
    (f_\external)\indices{^a}
    \partial\indices{_a} \phi
    ]^2
    \right]
    ,
\\
\label{eq:L_in_from_ex_general_with_vel}
    \mathcal{L}_\internal
&=
    \frac{1}{2} 
    \left[
    \eta
    \indices{^\mu^\nu}
    \partial\indices{_\mu} \phi
    \partial\indices{_\nu} \phi
    +
    \kcomp[\internal]^2
    \phi^2
    +
    \gamma_\sonic(c)^{-2}
    [(f_\internal)\indices{^\mu}
    \partial\indices{_\mu} \phi
    ]^2
    \right]
    .
\end{align}
Crucially, note that these forms are invariant under their respective Lorentz transformations.%
\footnote{Explicitly, let $x\indices{^{a'}} = (\Lambda_\sonic)\indices{^{a'}_a} x\indices{^a}$ and $x\indices{^{\mu'}} = \Lambda\indices{^{\mu'}_\mu} x\indices{^\mu}$ be a different set of coordinates, with $\Lambda_\sonic$ being a sonic Lorentz transformation and $\Lambda$ being an ordinary Lorentz transformation independent of $\Lambda_\sonic$. Then, the transformation between these coordinates is $x\indices{^{\mu'}} = K\indices{^{\mu'}_{a'}} x\indices{^{a'}}$, with
    $
    K\indices{^{\mu'}_{a'}}=
    (\Lambda_\sonic)\indices{_{a'}^a}
    \Lambda\indices{_\mu^{\mu'}} 
    K\indices{^\mu_a}
    $%
    .
The forms of Eqs.~\eqref{eq:L_ex_from_in_general_with_vel} and~\eqref{eq:L_in_from_ex_general_with_vel} are maintained in these new coordinates, with all indices now taking primes. In particular (and as already mentioned), $(f_\external)\indices{^a}$ and $(f_\internal)\indices{^\mu}$ behave as sonic and ordinary four-velocities, respectively, under these transformations.}

In fact, if we compare Eqs.~\eqref{eq:L_ex_from_in_general_with_vel} and~\eqref{eq:L_in_from_ex_general_with_vel},
we see that they are identical in form, both being special cases of
\begin{adjustbox}{max width=\columnwidth}
\parbox{\columnwidth}{%
\begin{align}
    \label{eq:L_part_from_obs_general_with_vel}
    \mathcal{L}_{\particle}^{(\observer)}
&=
    \frac{1}{2} 
    \left[
    (\eta_\observer)
    \indices{^\alpha^\beta}
    \partial\indices{_\alpha} \phi
    \partial\indices{_\beta} \phi
    +
    \kcomp[\particle]^2
    \phi^2
    +
    \gamma_\particle(c_\observer)^{-2}
    [(f_\particle)\indices{^\alpha}
    \partial\indices{_\alpha} \phi
    ]^2
    \right]
,
\nonumber
\\
\end{align}
}
\end{adjustbox}
where the subscripts `$\particle$' and `$\observer$', respectively, correspond to the particle and observer (internal or external for each), and the $\alpha\beta$ indices refer to observer coordinates. Thus, $\mathcal L_\particle^{(\observer)}$ represents the Lagrangian for particle type~$\particle$ as seen by observers of type~$\observer$. This form also accurately describes the case when the particle and observer are both of the same type, namely Eqs.~\eqref{eq:L_ex_from_ex}
and~\eqref{eq:L_in_from_in}, since $\gamma_\particle(c_\particle)^{-2} = 0$.
In this case, we abbreviate the notation as $\mathcal L_\particle \coloneqq \mathcal L_\particle^{(\particle)}$.

In what follows, we will explore the physical interpretation of Lagrangians of this form, paying particular attention to (a)~the generality of this framework and (b)~the physical interpretation of the three possible cases~$c_p < c_o$, $c_p > c_o$, and $c_p = c_o$.

\section{Internal and external observers on an equal footing}
\label{sec:observers}

It is illustrative to consider what happens if we introduce a second sonic medium. Suppose that the two sonic media have different densities, and hence different speeds of sound. We will refer to these media as `milk' and `honey', with respective LIS values of $c_\milk$ and $c_\honey$. For concreteness, assume that $c_\honey > c_\milk$. For each medium, we define {internal observers} (see Sec.~\ref{sec:sonics}) with coordinates indexed by `$\milk$' and `$\honey$', respectively. We also define 
ordinary ($c$-LIS) particles and 
a laboratory coordinate system, both 
denoted by~`$\mathcal \laboratory$'.

We limit our analysis to the case in which the media are mutually at rest. 
We choose this restriction because movement of the medium carrying excitations has been employed as a sonic model for gravitational effects~\cite{barceloAnalogueGravity2011}. We leave this possible extension of the current analysis to future work.
Here, we limit our discussion to the case of special relativistic effects in order to simplify the presentation and isolate the effects of observer motion and the varying LIS of the fields.

We consider one phonon field native to the milk and one native to the honey, as well as a laboratory-native scalar field. This gives us nine possible Lagrangians~$\mathcal{L}_\particle^{(\observer)}$, with $\particle, \observer \in \{\milk, \honey, \laboratory\}$. Seven of them will necessarily be
of the form of Eq.~\eqref{eq:L_part_from_obs_general_with_vel}:
\begin{align}
    \{
    \mathcal L_{\milk},
    \mathcal L_{\honey},
    \mathcal L_{\laboratory},
    \mathcal L_{\milk}^{(\laboratory)},
    \mathcal L_{\honey}^{(\laboratory)},
    \mathcal L_{\laboratory}^{(\milk)},
    \mathcal L_{\laboratory}^{(\honey)}
    \}
    .
\end{align}
We turn our attention instead to the remaining two, $\mathcal{L}_{\milk}^{(\honey)}$ and $\mathcal{L}_{\honey}^{(\milk)}$, since these represent how one {internal observer} would view the foreign particle.

To ground our calculations in straightforward laboratory physics, we will use the laboratory physics as a bridge, focusing on the behaviour of the milk and honey with respect to the laboratory---i.e.,~$\mathcal{L}_{\milk}^{(\laboratory)}$ and $\mathcal{L}_{\honey}^{(\laboratory)}$. We choose the following coordinate conventions:
\begin{align*}
    &\mu\,\nu & &= &&\text{laboratory coordinates},
\\
    &a\,b & &= &&\text{milk coordinates},
\\
    &j\,k & &= &&\text{honey coordinates}
    .
\end{align*}
We define the lab-to-milk transformation $K\indices{^a_\mu}$ and the lab-to-honey transformation $K\indices{^j_\mu}$ using Eqs.~\eqref{eq:xtrans} for each medium. Using the laboratory coordinates as a bridge, we now define the honey-to-milk transformation
\begin{align}
    K\indices{^a_j}
&\coloneqq
    K\indices{^a_\mu}
    \bar K\indices{^\mu_j}
    .
\end{align}
Notice that this transformation is unaffected by any Lorentz transformation on the laboratory coordinates.

The milk phonons as viewed from the laboratory coordinates are governed by Eq.~\eqref{eq:L_part_from_obs_general_with_vel}, 
\begin{align}
    \mathcal L_{\milk}^{(\laboratory)}
=    %
    \frac{1}{2} 
    \left[
    \eta
    \indices{^\mu^\nu}
    \partial\indices{_\mu} \phi
    \partial\indices{_\nu} \phi
    +
    \kcomp[\milk]^2
    \phi^2
    +
    (\eta_\milk - \eta)
    \indices{^\mu^\nu}
    \partial\indices{_\mu} \phi
    \partial\indices{_\nu} \phi
    \right]
.
\end{align}
Using prior steps in the derivation, we can write this in terms of an observer in the honey. When transforming from laboratory to honey coordinates contractions survive intact, and thus
\begin{align}
\label{eq:L_milk_from_honey}
&    \mathcal{L}_{\milk}^{(\honey)}
\\
&\;\nonumber =
    \frac{1}{2} 
    \left[
    (\eta_\honey)
    \indices{^j^k}
    \partial\indices{_j} \phi
    \partial\indices{_k} \phi
    +
    \kcomp[\milk]^2
    \phi^2
    +
    (\eta_\milk - \eta_\honey)
    \indices{^j^k}
    \partial\indices{_j} \phi
    \partial\indices{_k} \phi
    \right].
\end{align}
Because the media are mutually at rest, $\eta_\milk$ and $\eta_\honey$ are both diagonal in each others' coordinates (compare Eqs.~\eqref{eq:eta_inv_other_coords}), so we see that
\begin{align}
\label{eq:L_milk_from_honey_four_vector}
&    \mathcal{L}_{\milk}^{(\honey)}
\\
&\;\nonumber =
    \frac{1}{2} 
    \left[
    (\eta_\honey)
    \indices{^j^k}
    \partial\indices{_j} \phi
    \partial\indices{_k} \phi
    +
    \kcomp[\milk]^2
    \phi^2
    +
    \gamma_\milk(c_\honey)^{-2}
    [(f_\milk)\indices{^j}
    \partial\indices{_j} \phi
    ]^2
    \right].
\end{align}
By an entirely analogous argument, we have the following result for the honey phonons as viewed by the milk observers:
\begin{align}
\label{eq:L_honey_from_milk_with_vel}
&    \mathcal{L}_{\honey}^{(\milk)}
\\
&\;\nonumber =
    \frac{1}{2} 
    \left[
    (\eta_\milk)
    \indices{^a^b}
    \partial\indices{_a} \phi
    \partial\indices{_b} \phi
    +
    \kcomp[\honey]^2
    \phi^2
    +
    \gamma_\honey(c_\milk)^{-2}
    [(f_\honey)\indices{^a}
    \partial\indices{_a} \phi
    ]^2
    \right].
\end{align}
The end result is that all nine Lagrangians $\mathcal{L}_\particle^{(\observer)}$, for each pair of $\particle, \observer \in \{\milk, \honey, \laboratory\}$, are of the form of Eq.~\eqref{eq:L_part_from_obs_general_with_vel}.

Of course, there is nothing special about the milk or the honey. They are both sonic media and so the scenario is symmetric---any observer can believe
that their relativity is valid (i.e.,~that the `true' symmetries of nature are the Lorentz symmetries with the invariant speed of \emph{their} signals), and that all other signals are propagations in a medium with a preferred rest frame. There is a kind of `relativity of relativities' at work here, of a kind stronger than the one that arises when one considers a `sonic' experiment within a `photonic' lab, as in~\cite{TM}.\footnote{Indeed, we take this to be exactly a case of the `conventionality of geometry', famously associated with Poincar\'{e}~\cite{Poincare} and Reichenbach~\cite{Reichenbach} (for a review, see~\cite[\S 4]{PhilCompass}).} The only possibly meaningful distinguishing feature lies in the sign of the last term, which determines whether the medium carrying the particle appears as an ordinary acoustic medium to the observer ($c_o > c_p$) or a novel \emph{tachyonic} medium ($c_o < c_p$), a concept that will be properly introduced in Sec.~\ref{sec:relofrel}.

The argument above illustrates 
the point 
that an observer in a sonic medium %
will not determine that they exist within a medium.
Any signals propagating outside the medium will be described by a Lagrangian that is equivalent to that for particles internal to the medium, plus a correction term that looks like a Lorentz violation. An observer can freely ascribe any Lorentz-violating effects to the foreign particles.
No {internal observer} has any reason to construct a concept of Lorentz invariance that pays attention to anything other than their own local LIS, and hence all observers (including those in the laboratory, basing their concept of Lorentz invariance on photons) are on an equal footing. 

We emphasize that %
we are considering {internal observers}, as discussed in Sec.~\ref{sec:intro}. These observers are compelled to base their length and time scales on the exchange of sound signals, and are then acting rationally to conclude that sound signals propagate isotropically, apply Occam's razor, and conclude that they do not exist within a medium at all. (For further discussion of this, see~\cite{ChengRead,TM}.)

One radical conclusion implied by the universality of the Lagrangian Eq.~\eqref{eq:L_part_from_obs_general_with_vel} for both external and internal observers
is that
an internal observer will be able to argue (by setting $c_p = c$) that their own sound signals propagate free of any medium and furthermore that the light-based relativity in the lab violates (acoustic) Lorentz invariance. Hence, to this observer, light signals have an implied rest frame and a (presumed) medium through which they propagate. (Of course, this does not say anything about the physical nature of that medium.)

This seems to violate relativity as per Einstein, but in fact it does not because observers cannot measure the one-way LIS. Keeping in mind the commonly-used image of a reference frame as a series of correlated clocks and rulers, it might be that light (or sound, for {internal observer}s) moves faster in one direction than the opposite direction. This, however, cannot be determined by an observer who uses the exchange of those same signals to synchronize their clocks and hence determine the lengths of their rulers. (For a detailed discussion of this, see~\cite{Salmon1977-SALTPS}.) It follows that the structure of special relativity is a consequence of choosing a particular type of signal to synchronize clocks, and \emph{postulating by fiat} that such signals travel isotropically at a special speed $c_o$. The choice of special speed is a conventional one (rather than being entirely prescriptive) that is informed by the types of signals available to be sent and received. While the choice of this speed is, at its core, arbitrary, some choices lead to simpler theories than others. We discuss this further in the next section.

\section{Relativity of relativities}\label{sec:relofrel}

\begin{figure}[t]
    \centering
  \includegraphics[width=\linewidth]{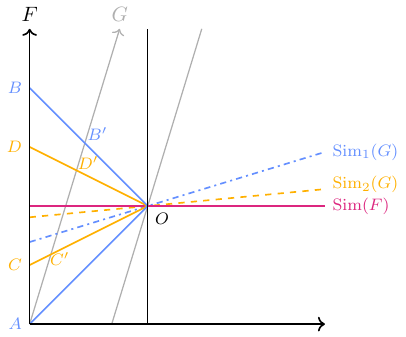}
  \caption{\label{fig:clocks}Two Langevin clocks $F$ and $G$ moving uniformly with respect to one another, with two different bouncing signals with different two-way speeds (forming triangles $AOB$ and $COD$). The `standard' synchrony conventions for each of the two signals agree in the rest frame of $F$, but not in the rest frame of $G$, as is clear from the fact that $\text{Sim}_1(G) \neq \text{Sim}_2(G)$. Thus, choosing a simultaneity convention that renders one signal isotropic in all inertial frames of necessity renders all others generically anisotropic (except in one frame: here, the rest frame of $F$), and vice versa.}
  \vspace{-3mm}
\end{figure}

Let us expand on this further. Consider the setup shown in Figure~\ref{fig:clocks}. Here, we have two Langevin clocks (i.e.,~clocks consisting of two mirrors and a bouncing signal), $F$ and $G$, moving uniformly with respect to one another; the setup is considered in the rest frame of $F$, so the two mirrors of $F$ are represented by the two black vertical lines, whereas the two mirrors of $G$ are represented by the two grey diagonal lines.

First, consider the bouncing signal forming the triangle $AOB$ ($\bigtriangleup{AOB}$). One can ask: which point on the worldline of the first mirror is simultaneous with the `bounce' event on the second mirror? If one adopts the distant clock synchrony convention 
proposed
by Einstein in his 1905 article---now known as \emph{Einstein-Poicar\'{e} synchrony} or \emph{standard synchrony}---then one will stipulate that the event half-way between the events of emission and reception on the first mirror will be simultaneous with the bounce event; thus, one will draw a horizontal simultaneity surface $\text{Sim}(F)$, represented by the horizontal solid line.

Next, consider 
the rest frame of $G$ in motion with respect to~$F$. 
Using the same signal defining $\bigtriangleup{AOB}$, one can ask: which point on the worldline of the first mirror of $G$ is simultaneous with the `bounce' event? If one again avails oneself of standard synchrony, one will now draw the dash-dotted simultaneity hypersurface $\text{Sim}_1(G)$ (bisecting the line segment $\overline{AB^\prime}$ and passing through the point $O$), tilted with respect to $\text{Sim}(F)$. This is nothing but the familiar relativity of simultaneity: choosing standard synchrony in the rest frames of both $F$ and $G$ causes simultaneity hypersurfaces to tilt as one views the situation from different rest frames%
, and this has the merit of rendering the one-way velocity of the signal isotropic in \emph{any} frame moving uniformly with respect to~%
$F$.

But now introduce a second signal, faster than the first, forming the triangle $COD$ ($\bigtriangleup{COD}$) in  Fig.~\ref{fig:clocks}. If one uses this signal to synchronize clocks in the rest frame of $F$, then one again produces the simultaneity hypersurface $\text{Sim}(F)$ and renders the one-way velocity of this signal isotropic.

What about when one applies standard synchrony to this signal in the rest frame of $G$? In this case, one no longer produces the simultaneity hypersurface $\text{Sim}_1(G)$ but rather the distinct simultaneity hypersurface $\text{Sim}_2(G)$ (bisecting the line segment $\overline{{C^\prime}{D^\prime}}$ and passing through the point $O$), represented by the dashed line. The latter simultaneity hypersurfaces---those formed with respect to $\bigtriangleup{COD}$---are \emph{not} those picked out by standard synchrony with respect to $\bigtriangleup{AOB}$---that is, they do not render the one-way speed of the signal forming $\bigtriangleup{AOB}$  isotropic in the rest frame of $G$ (but they \emph{do} achieve this for the signal forming $\bigtriangleup{COD}$). Likewise, the former simultaneity hypersurfaces---those formed with respect to $\bigtriangleup{AOB}$---are not those picked out by standard synchrony with respect to $\bigtriangleup{COD}$---that is, they do not render the one-way speed of the signal forming $\bigtriangleup{COD}$ isotropic.

Here, then, is the rub. Special relativity \emph{\`{a} la} Einstein 1905 consists of several inputs, among them the relativity principle, the light postulate, and an assumption of standard synchrony. But with multiple signals, one has a conventional choice as to which such signal to apply standard synchrony with respect to: having chosen one such signal (say the signal forming $\bigtriangleup{AOB}$ in Figure~\ref{fig:clocks}), then in all but one frame, the one-way speed of \emph{other} signals will be rendered anisotropic. Thus, insofar as there is a conventional choice as to which signal to use in adjudications of the simultaneity of distant events, there is a conventional chose as to which signals propagate isotropically and which do not. (These connections are also noted at e.g.~\cite[p.~172]{Liberati}.)

In this sense, then, there is no absolute standard of isotropy. Signals can only be isotropic or anisotropic relative to each other. This even goes for signals that travel faster than the special speed. %
Ultimately, what breaks the symmetry in a choice between two signal speeds can only be \emph{dynamical} considerations. For example, from the laboratory standpoint, the standard model of particle physics takes a particularly simple form when $c$ is chosen as the LIS for special relativity. This choice happens to be consistent with the empirical observation that nothing can travel faster than~$c$---i.e.,~this choice of LIS also happens to serve as an ultimate speed limit---but the reader should take care to note that such an observation does not \emph{compel} the choice of $c$ as the LIS. One remains free, instead, to choose (say) the speed of sound as the LIS for presenting the standard model. The impediment to doing so is much more complicated dynamics and failure to reveal the simplicity of the laws and symmetries that are present when one chooses $c$~as the LIS.

For example, if one were to construct physical laws based around a speed $c_\sonic < c$, such that the Lorentz factor were $1/\sqrt{1-v^2/{c_\sonic}^2}$, not only would the measured mass of an object increase as it approached $c_\sonic$, but it would become an imaginary quantity as the object exceeded $c_\sonic$. To account for collisions between tardyonic and tachyonic objects, the mathematical formulation of conservation of momentum would have to be written in a manner that took both real and imaginary momenta into account. This would be possible, but it would be more cumbersome than simply recognising that light has the fastest speed of any signal we know of and choosing $c$ as the speed of isotropic propagation. 

We have already seen that for internal observers in a sonic medium, this logic does not hold due to the imposed restrictions on their ability to interact with particles that do not \emph{dynamically} reflect the sonic relativity of their world. As such, they are right to choose a speed other than~$c$ (specifically, $c_\sonic$) as their LIS since it simplifies their description of the world to which they have access. The ubiquity of this fundamental---and often overlooked---freedom in the description of physical laws is discussed further in Ref.~\cite[ch.~2]{brown2005physical}.

Note that the choice of a particular clock synchrony convention is just the choice of a particular coordinatisation of space and time---and it is an utterly prosaic fact of physics practice that some coordinate systems lead to simpler descriptions of physical processes than others. What we are stressing here is that light is not, in some sense, privileged over sound at the level of \emph{kinematics} (perhaps contrary to the thinking of the later Einstein---we return to this below). Rather, the symmetry between the two is broken at the level of \emph{dynamics}, and in particular at the level of a pragmatic/conventional choice as to what simplifies the dynamics maximally. Nevertheless, just as one is free to choose to coordinatize problems in physics in any way in which one pleases, in principle one is free to select \emph{any} signal as being the `special' signal.

Here is another way to make the point. In Ref.~\cite{ChengRead}, the authors ask the following question: ``What would be wrong with replacing the light postulate in Einstein's 1905 article on special relativity with a `sound postulate'?'' The answer that the authors offer has to do with the fact that there is ample empirical evidence for a medium for sound (so goes an extremely standard line of reasoning, which is endorsed in Ref.~\cite{ChengRead}), but no such evidence for a luminiferous ether. What we wish to stress here is that one could---in principle---stick to one's guns by applying standard synchrony with respect to sound signals, thereby rendering the one-way speed of sound isotropic in all frames moving uniformly with respect to one another, and thereby rendering the one-way speed of all other signals (including light!) \emph{anisotropic} in all but one frame---in other words, affording those signals a preferred frame, which could be identified with the rest frame of their medium. It should be recognized that such a medium need not actually exist. However, the \emph{presumed} existence of a medium (or media) will be consistent with the mathematical description arising from any given choice of isotropic signal.  Practical considerations might weigh against this decision---but there is no \emph{a priori} prohibition against it.

\section{The prospect of retrocausal signalling}\label{sec:retrocausal}

Since the speed of sound can be different in different media, a question that arises naturally is whether we have created a sonic model of `faster-than-light' signalling between internal observers by allowing them to send signals faster than their own LIS using the particles of a different, faster medium---or by using external particles (which are not bound to a medium). We call this scenario \emph{tachyonic signalling}.%
\footnote{While the term as defined here applies only to internal observers sending supersonic signals, and our physical claims are limited to this case, we intentionally chose a term that would also apply to (the hypothetical possibility of) superluminal signalling by external observers.}
Due to the appearance of a foreign medium%
\footnote{This perspective holds for the internal observers regardless of whether the apparent medium is a real object or not (from the laboratory perspective), as discussed in Sec.~\ref{sec:relofrel}.}
having a propagation speed that is faster than that at which the observers' excitations propagate, we dub this a `tachyonic
medium'. (The idea of a tachyonic medium was also considered briefly in Ref.~\cite[\S 5.4]{ChengRead}.) This is in contrast to an acoustic medium---a mundane concept with which all readers will be familiar---that carries excitations that propagate slower than the observers' excitations.

Besides the fact that it has never been observed, a long-standing objection to the prospect of tachyonic signalling is that it would allow an observer to send information into their own past light cone, thereby violating causality (there is a large literature on this topic: see e.g.,~Ref.~\cite{Arntzenius} and references therein for discussion). The acoustic models described above allow us to investigate the consequences of such signalling if it were possible. (Geroch~\cite{Geroch}, for example, suggests that such signalling is possible, but does not explore the idea further.)  

\begin{figure}[t]
    \centering
  \includegraphics[width=\linewidth]{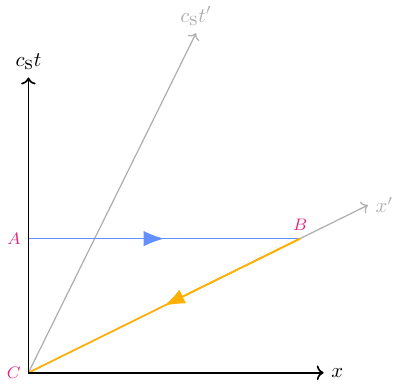}
  \caption{\label{fig:round-trip01}The standard scenario by which tachyonic signalling is argued to admit causality violation~\cite{Benford} involves two observers in relative motion. If each observer can send a signal that they perceive as moving sufficiently fast (in this example, infinitely fast, so that each signal travels parallel to the emitter observer's $x$ axis), it is possible for information to perform a round trip into the past light cone of either observer. In this example, the first observer (non-primed coordinates) emits a tachyonic signal at event $A$ in the $+x$ direction. A second observer (primed coordinates), in motion away from the first, receives it at event $B$ and transmits a return signal in the $-x'$ direction. This signal arrives at event $C$, which lies in the past light cone of event $A$, thus violating causality.}
\end{figure}

In the context of this paper, any signal that moves faster than the LIS of the observer is referred to as `tachyonic'.
Taking a cue from science fiction, we refer to a device that permits tachyonic signalling as a `subspace communicator'.%
\footnote{Several other names have been coined for fictional tachyonic communication devices including `interocitor', `Dirac communicator', and `ansible'.}
We are adopting the convention whereby `subspace' is a distinct domain %
that admits tachyonic transmission of signals. %
Hence, it is entirely reasonable for observers in one acoustic toy universe to regard another acoustic toy universe with a higher LIS as a subspace. %
There is no reason to presume that a signal cannot be passed between the `slow' and `fast' universes.
Suppose we have two observers, Marty and Emmett, equipped with subspace communicators. %
It is generally held that Emmett may send information into his own past light-cone, as we can see by reference to the Minkowski diagram in Fig.~\ref{fig:round-trip01}. For simplicity we can assume that tachyonic communication is, effectively, instantaneous, and hence signals travel parallel to the transmitting observer's spatial axis. At event~$A$ Emmett sends a message to Marty, who is moving away from Emmett and receives the signal at event~$B$. Marty then immediately re-transmits the signal, which travels along his $-x$ axis back to Emmett who receives it at event~$C$. Owing to the angle between the spatial axes of Emmett and Marty, the event~$C$ occurs before event~$A$ (Fig.~\ref{fig:round-trip01}). Such an arrangement of subspace communicators that would permit an observer to send a message into their own past light-cone is sometimes called a `tachyonic anti-telephone' (Ref.~\cite{Benford}). %
This scenario violates causality, as event~$C$ may trigger a process that prevents the signal at $A$ from being emitted, potentially leading to a version of the so-called `grandfather paradox'.

This leads to three basic responses. The first is to contend that tachyonic signalling is impossible. The second is to invoke a chronology protection conjecture~\cite{hawking1992chronology}, implying that the laws of physics prevent the formation of time machines (the thought being that, e.g.,~quantum mechanical effects will destabilize any spacetime geometry that would allow retrocausal signalling). The third is to invoke some variant of the Novikov self-consistency principle, which argues that only self-consistent time travel histories are `stable'. (In the foundations of relativity, these options are discussed further in, e.g., Ref.~\cite{Earman}.)
We present in the next section an explicit resolution to this problem in our acoustic model, which establishes a chronology protection conjecture that is entirely classical.

\section{Tachyonic media and the preservation of causality%
}\label{sec:tachyonicmedia}
In the acoustic model of tachyonic signalling that we have described, any signal (such as a light signal moving in the lab) will respect causality, so it appears that we should be unable to build an acoustic version of the tachyonic anti-telephone (Fig.~\ref{fig:round-trip01}). Breaking the assumption of isotropic tachyonic signalling is the key. As we have seen in Sec.~\ref{sec:relofrel}, once one picks a type of signal to be isotropic, there can be only one---all the others have to adjust accordingly. In the case under consideration here, an observer will perceive any tachyonic signal to be travelling as though it is moving through an ether with a preferred rest frame. If Emmett and Marty are in relative motion, they cannot both be at rest relative to the tachyonic signal's medium, and the resulting anisotropy of tachyonic signals prevents retrocausal signalling since a round-trip tachyonic signal always propagates forward in time with respect to the tachyonic medium.

\begin{figure}[t]
    \centering
  \includegraphics[width=\linewidth]{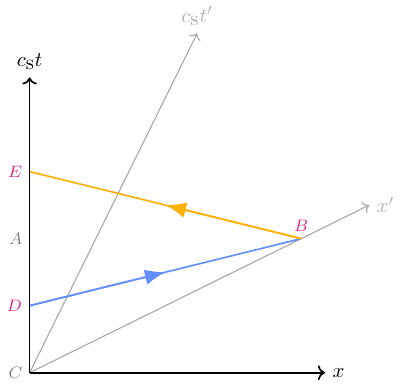}
  \caption{\label{fig:round-trip02}The apparent presence of a tachyonic medium is consistent with preservation of causality. In this figure, the tachyonic medium is at rest with respect to the non-primed axes, and tachyonic signals always propagate isotropically with respect to the non-primed coordinates. When an observer tries to send information via tachyonic signalling into their own past by following the procedure shown in Fig.~\ref{fig:round-trip01} (here shown using a finite---but sufficiently fast---signal speed for clarity), the existence of a rest frame for the tachyonic signals ensures the return signal always has a four-velocity with a positive $t$ component. Thus, tachyons confined to a medium cannot be used to create a tachyonic anti-telephone.%
  }
\end{figure}

To see how this prevents retrocausal signalling, consider Fig.~\ref{fig:round-trip02}, with Emmett using the non-primed axes and Marty the primed axes. We consider the case in which the tachyonic medium is at rest with respect to Emmett. Thus, if Emmett and Marty try to use excitations of this tachyonic medium to send a round-trip signal as in Fig.~\ref{fig:round-trip01}, the outbound signal (now shown with a finite but sufficiently fast signal speed) will follow the path~$D$ to~$B$, and the return signal will follow the path $B$ to~$E$, with the return signal arriving at Emmett (at~$E$, just after~$A$) after he has sent the outbound signal (from~$D$, just before~$A$), as should be the case.

In essence, the apparent existence of a tachyonic medium makes Marty's motion relative to Emmett unimportant, as the path of the tachyonic signal through spacetime is determined by the tachyonic medium, not the orientation of Marty's spatial axis. As the speed of the signal relative to Emmett approaches infinity, the temporal separation between events $D$ and $E$ becomes arbitrarily small---but never reverses sign.

\begin{figure*}[t!]
    \centering
    \includegraphics[width=\linewidth]{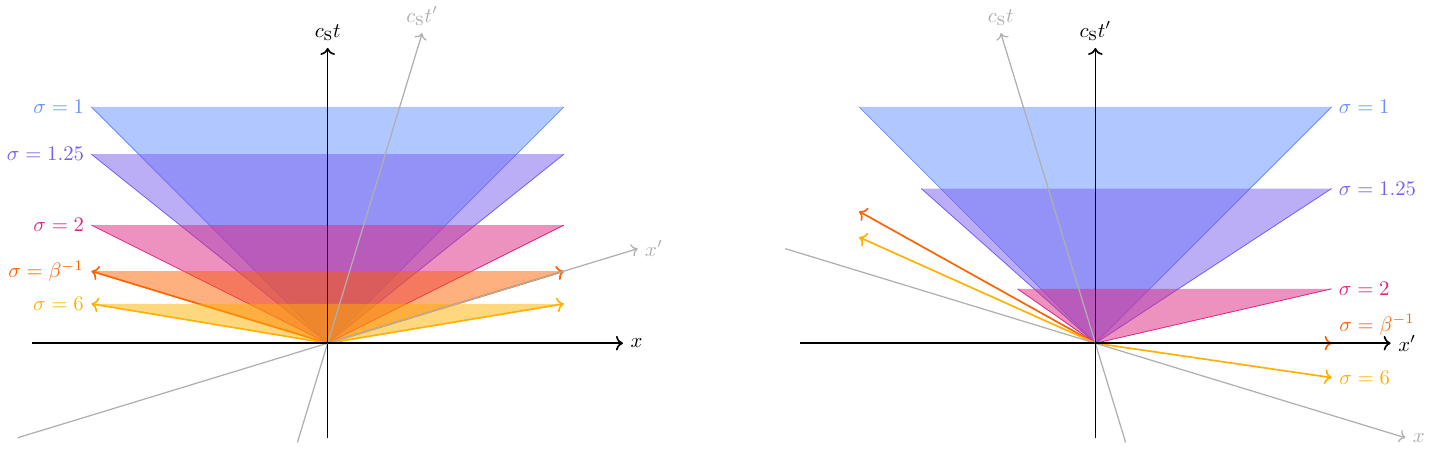}
    \caption{The propagation of signals in the rest frame of the tachyonic medium 
    (left) and in 
    a frame moving with dimensionless velocity $\beta=\sqrt{41}/21\approx{0.3049}$ with respect to the medium (right), as seen by internal observers. In both cases, the other coordinate system is shown in grey. Signal velocities $\sigma$ (the ratio of the signal speed to the speed of sound $c_\textrm{s}$ in the medium) define `signal cones' whose upper boundaries correspond to constant time values in the frame of interest. Signals having $\sigma=1$ follow null curves, defining a `sound cone' that is invariant under sonically relativistic Lorentz transformations. Only in the tachyonic medium's frame do signals with
    $\sigma\neq 1$ propagate isotropically. The same signal cones in the moving frame are tilted, indicating anisotropic propagation. (Still, tachyonic signals remain tachyonic for all observers since they travel along spacelike intervals.) For $\sigma=\beta^{-1}$ in the medium, moving observers measure an infinite signal velocity (i.e., parallel to the $x^\prime$ axis) in the $+x^\prime$ direction, and a finite velocity (i.e., making a non-zero angle with the $x^\prime$ axis) in the $-x^\prime$ direction. Furthermore, $\sigma>\beta^{-1}$ signals propagating in the positive spatial direction appear to travel into the \emph{past}, while still propogating into the future in the negative spatial direction. This asymmetry ensures an observer can never send a signal into their own past.}
    \label{fig:Tachyonic_signal_Cones_All_In_One}
\end{figure*}

Notice that there is nothing preventing a tachyonic signal being sent from Emmett when his clock shows time~$T$ and arriving at Marty when his clock shows a time earlier than~$T$. It is not possible, however, for Marty's return signal to arrive back at Emmett before his own clock reaches~$T$. Marty's return signal always arrives at Emmett's location as or after he sends it, never before, and this holds true in any reference frame. Figure~\ref{fig:Tachyonic_signal_Cones_All_In_One} illustrates the anisotropy of tachyonic signals for a moving observer.

There should be nothing surprising about the anisotropic propagation of tachyonic signals. Most signals propagate anisotropically, relative to observers outside the medium through which they propagate. Light is just unusual because it does propagate isotropically (as far as we are concerned---and ultimately, as we have seen, by stipulation). In short, the only significant difference is that sound waves are anisotropic and slow; tachyons (in this model) are anisotropic and fast.

That causality is preserved by the anisotropy of tachyonic signalling fits in well with the `schematic' of causality-respecting superluminal signalling offered by Carballo-Rubio et al.~in Ref.~\cite[\S 10]{carballo-rubioCausalHierarchyModified2020}. There, the authors make the case that the conventional logic that superluminal (read: tachyonic) communication within relativistic theories implies causality violation can be turned around, and thus that a causality-respecting relativistic theory incorporating superluminal (read: tachyonic) communication can be obtained at the expense of some extra structure within the theory. In our model, the extra structure that saves causality for internal observers is that of a preferred rest frame \textit{for the tachyonic signals}.

The work of Liberati et al.~\cite{Liberati} also makes a point similar to that of Carballo-Rubio et al.,~stating ``As far as causality is concerned, it is impossible to make statements of general validity, without specifying at least some features of the tachyonic propagation.'' In essence, whether or not causality is respected or broken is undecidable until extra structure (``...at least some features of the tachyonic propagation.'') has been provided. In particular, Liberati et al.~\cite{Liberati} come to a similar conclusion as the current work regarding the extra structure required to avoid casual paradoxes with tachyonic signalling, stating ``Obviously, there can be no paradox if, in one particular reference frame, tachyons can only propagate forward in time.'' This is true within the rest frame of the tachyonic medium for the model that we have discussed within this current work.

\section{Conclusions}\label{sec:conclusion}

Our discussion was initially framed in the context of an `external' {domain} (the laboratory) containing an `internal' sonic {domain} (the medium)---in which case it is tempting to think of the external {domain} as being truly Lorentz obeying and the internal one as being truly Lorentz violating. In particular, observers may determine that signal speeds in another {domain} can exceed those in their own. This may be interpreted as a hint that their {domain} exists in a medium within a larger `external' one.

As emphasized, however, given that Eq.~\eqref{eq:L_part_from_obs_general_with_vel} does not pick out a preferred {domain}, the observers in either {domain} may deduce that Lorentz symmetry is obeyed in their own (since it would be preserved by every other experiment with which the observer is familiar) and that the other {domain} is endowed with a preferred rest frame. We must recognize that this deduction is perfectly reasonable in either one
(i.e.,~it comes about as a result of the observers applying Occam's razor). 
It is one particular case of the Poincar\'{e}-Reichenbach conventionality of geometry: various different geometrical descriptions of physical goings-on are possible; only super-empirical considerations of simplicity, etc.,\ can decide between these (Ref.~\cite{PhilCompass}).

What is novel in this formulation is that we have a physical model that would naturally lead rational observers to posit a medium that carries tachyonic signals rather than acoustic ones if, for instance, some intelligent {internal observer}s discovered light. (Or---hypothetically---if intelligent external observers discovered superluminal particles.) We, as external observers, have our entire theory of relativity founded upon the speed of light and thus consider the phonons used by the {internal observer}s to construct clocks and rulers to be travelling through an acoustic medium. This is right and proper since it is the starting point for the acoustic-observer toy model. The {internal observer}s, however, cannot ever be proven wrong when they believe that (1)~their relativity is based upon the speed of sound and is observably legitimate, and (2)~they interpret \emph{photons} as excitations in a tachyonic medium that has a definite rest frame.

This is surprising, yet there is nothing that can be done, under the assumptions made by {internal observer}s, to convince them
that the existence of (supersonic) photons means that it is the observers' precious \emph{phonons} that are the medium-bound excitations, while the \emph{photons} are medium free---which is, of course, what we believe from the outside. Furthermore, if causality in the external {domain} is preserved, we expect it to also be preserved for the {internal observer}s. 

The apparent existence of a preferred rest frame for the external {domain} ensures that the order of events observed in %
either {domain} must be the same. This drives home the fact that the choice of which {domain} is `truly' fundamental is arbitrary and conventional as far as the observers in either {domain} are concerned. Preservation of causality 
prohibits time-travel paradoxes, but in doing so, it removes a possible standard by which one choice of special speed may be judged preferable to another. The same reasoning could be applied 
outside of the analogue toy model presented here. Specifically, the
hypothetical future detection of tachyonic signalling would not necessarily invalidate our current understanding of physics---especially if observations of such signals were consistent with their propagation through a tachyonic medium.

We may, in fact, invert the reasoning above to conclude that the structure of special relativity arises from a fundamental symmetry of nature:~the freedom to choose which speed corresponds to isotropic (medium-free) propagation of information. This symmetry is broken by the choice to synchronize clocks and rulers using the associated signals (light pulses, sound pulses, etc.),\ and once this choice is made, all other signals, whether faster or slower, appear to propagate anisotropically. It is this freedom to choose a special speed---rather than some ad-hoc prohibition against tachyonic signaling---that ensures that causality is preserved.     

We will close by pointing out an irony in Einstein's own writings on special relativity. Later in his life, Einstein wrote:
\begin{quote}
The special theory of relativity grew out of the Maxwell electromagnetic equations. But\ldots the Lorentz transformation, the real basis of special-relativity theory, in itself has nothing to do with the Maxwell theory.~\cite{Einstein1935}
\end{quote}
That is to say, Einstein would later come to regret the apparent special role of light in his 1905 article. Rather, in the view of the later Einstein, the situation was this:
\begin{quote}
The content of the restricted relativity theory can accordingly be summarized in one sentence: all natural laws must be so conditioned that they are covariant with respect to Lorentz transformations.~\cite{Einstein1954}
\end{quote}
That is, the later Einstein insists that special relativity is best understood as a universal \emph{kinematical} constraint of Lorentz invariance---presumably, with invariant speed~$c$ (this is now a very common view on the theory). In one sense, this clearly makes no (explicit) mention of light. But on the other hand, the implicit choice of $c$ continues to locate
the special theory within an electromagnetic paradigm. %

None of this is to say that Einstein was wrong. He was right to deduce that lengths and durations are not absolute---and spectacularly so! One can, indeed, using the machinery of the 1905 article, select \emph{any} signal to be the `special' signal, apply standard synchrony (one of the conventional inputs of the 1905 approach) with respect to that signal, and deduce the mathematical structure of special relativity. As discussed, this would impact the simplicity of the dynamical laws, but at the kinematical level, this is not a consideration.

It is ironic that the later Einstein viewed his universal kinematical constraint as liberating special relativity from electromagnetism. We would suggest, to the contrary, that the opposite is the case. Precisely what is to be resisted, in fact, is the view that $c$ has any special kinematical significance over other signal speeds. %
\vspace{3ex}

\acknowledgments

We thank Harvey Brown for fascinating and valuable discussions. This work was supported the Australian Research Council through a Discovery Project (Project No.\ DP200102152) and through the Centre of Excellence for Quantum Computation and Communication Technology (Project No.\ CE170100012).
J.R.~acknowledges the support of the Leverhulme Trust.

\bibliography{refs}%

\appendix

\end{document}